\documentclass[final,1p,times]{elsarticle} 
\usepackage{graphicx} 
\usepackage{amssymb} 
\usepackage{amsthm} 
\usepackage{lineno} 

\newcommand{\pt}{$p_{\rm T}$}                    
\newcommand{\et}{$E_{\rm T}$}

\journal{Nuclear Physics A} 
\begin{document} 

\begin{frontmatter} 


\title{Triggering on Hard Probes in Heavy-Ion Collisions with the CMS Experiment at the LHC}

\author{Christof Roland for the CMS Collaboration}

\address{MIT, Cambridge, USA.}

\begin{abstract} 
Studies of heavy-ion collisions at the LHC will benefit from an array of qualitatively new probes not readily available at lower collision energies. These include fully formed jets at $E_T > 50$~GeV, Z$^0$'s and abundantly produced heavy flavors. For Pb+Pb running at LHC design luminosity, the collision rate in the CMS interaction region will exceed the available bandwidth to store data by several orders of magnitude. Therefore an efficient trigger strategy is needed to select the few percent of the incoming events containing the most interesting signatures. In this report, we will present the heavy-ion trigger strategy developed for the unique two-layer trigger system of the CMS experiment which consists of a ``Level-1'' trigger based on custom electronics and a High Level Trigger (HLT) implemented using a large cluster of commodity computers.

\end{abstract} 

\end{frontmatter} 


\section{Introduction}\label{intro}

Heavy-ion collisions at the Large Hadron Collider (LHC) will provide a unique opportunity to study QCD matter at very high temperature.
Results from the experiments at the Relativistic Heavy Ion Collider (RHIC) provide 
insight into what can be expected at the LHC. They suggest that in heavy-ion collisions at 
200 GeV 
an equilibrated, strongly-coupled partonic system is formed.
There is strong evidence that this dense partonic medium is highly interactive, perhaps best described as a quark-gluon liquid, and is also almost opaque to fast 
partons. Measurements at the LHC will provide new, quantitative information about the nature and properties of this medium, by extending existing studies to much higher energy density and temperature and also by bringing to bear a broad range of novel probes. 

These probes include high \pt~ jets and photons, the $\Upsilon$ states, abundant D and B bosons and high-mass dileptons. 
The Compact Muon Solenoid Detector (CMS) \cite{cmsptdrv1} provides unique capabilities for detailed measurements that exploit these new opportunities at the LHC and will directly address the fundamental questions in the field of high density QCD. The key component in exploiting the CMS capabilities in heavy-ion collisions is the trigger system, which is crucial in accessing the rare probes expected to yield the most direct insights into the properties of high density strongly interacting matter.

\section{The CMS Trigger System}\label{trigger}

The unique CMS trigger architecture employs only two trigger levels: The Level-1 trigger is 
implemented using custom electronics and inspects events at the full bunch crossing rate. 
The Level-1 trigger uses local data from the calorimeter and muon systems to make
electron/photon, jet and energy sum, and muon triggers. 
The decision is sent to the front-end detector electronics after a latency of 3~$\mu$s. 
Events selected by the Level-1 trigger are then transferred to the High Level Trigger (HLT). 

All further online selection is performed in the HLT using a large 
cluster of commodity workstations (the ``filter farm'') running trigger algorithms on 
fully assembled event information. The HLT software environment allows the execution of 
complex ``offline'' analysis algorithms, restricted only by the execution time of these 
algorithms. 

\begin{figure}[ht]
\centering
\includegraphics[width=0.85\textwidth]{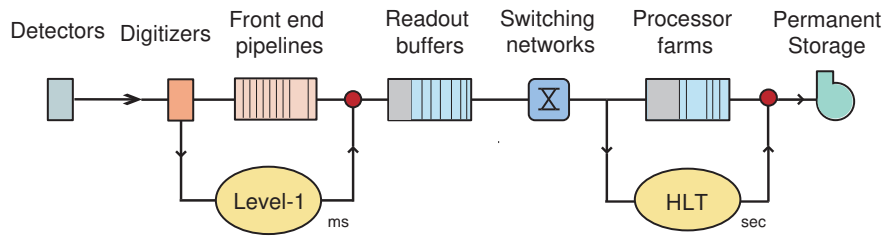}
\caption{Schematic of the data flow through the CMS Trigger system.}
\label{fig:data_flow}
\end{figure}

Events selected by the HLT are subsequently transferred to mass storage.
Figure~\ref{fig:data_flow} illustrates the 
data flow through the CMS trigger system.

\subsection{Trigger Strategy for Heavy Ion Collisions}

At Pb+Pb design luminosity, the initial collision rate at the beginning of a fill of Pb ions is 
expected to be close to 8~kHz. The luminosity evolution through a fill results in an average 
collision rate of about 3~kHz. 

In heavy ion running the main purpose of the CMS Level-1 trigger is to select true 
heavy ion collisions and trigger the detector readout while discriminating against 
beam gas interactions and non-collision related backgrounds. 
No significant rejection of Pb+Pb collisions at the Level-1 trigger is foreseen, but
trigger information from the Level-1 processing of the calorimeters and muon chambers 
can provide seed objects used to select the execution of specific HLT algorithms.

Every Pb+Pb collision identified by the Level-1 trigger will be sent to the HLT filter farm. 
At the HLT, the full detector information will be available for each event. All rejection
of Pb+Pb collisions required to reduce the rate of sampled events, i.e. the interaction rate, to 
the rate of events that can be transfered to mass storage will be based on the outcome of 
HLT trigger algorithms. For heavy ion running currently four trigger selections are foreseen:
\begin{itemize}
\item Minimum bias trigger. This event selection is based on the collision definition derived from the calorimeter information processed at Level-1. No further event selection is applied in the HLT.
\item Jet trigger. Jets are reconstructed by processing the calorimeter data with an iterative cone type algorithm including underlying event subtraction \cite{Kodolova2007}. Events can be selected based on the number and \et~ of the jets reconstructed in the event. 
\item Photon trigger. Photons are reconstructed by subjecting the electromagnetic calorimeter (ECAL) information to a clustering algorithm. Events are selected based on the presence of high \et~ ECAL clusters.
\item Dimuon trigger. For events containing two Level-1 muon candidates the primary event vertex is reconstructed and track segments in the muon chambers are propagated back through the CMS Si tracker to the event vertex. The event selection is based on fully reconstructed ``offline-quality'' dimuons. 
\end{itemize}

In order to adjust the data volume of the selected events to the available bandwidth to mass storage independent pre-scale factors are assigned to each selection. Based on the trigger objects reconstructed in the HLT varying pre-scale factors can be applied based on different \et~ thresholds of the same type of trigger object. Abundant low \et~ triggers can be recorded with an appropriate pre-scale factor while high \et~ triggers are recorded unprescaled, to make maximum use of the large cross section for high \et~ probes at the LHC.

\subsection {Trigger performance studies}

The purpose of the heavy ion trigger is the allocation of the available output bandwidth to a 
selection of trigger channels, such that the overall physics impact of the CMS heavy ion program
is maximized. This strategy requires that the algorithms can be executed quickly enough. Also, to 
appropriately distribute the available bandwidth to mass storage, the average data volume of 
the event selected by the various trigger selections has to be evaluated. Triggering 
on hard probes biases the selected event size towards larger central events owing to the binary 
collision scaling of the production cross section. 

To evaluate the performance of the heavy ion trigger strategy a data sample of 
35k minimum bias Pb+Pb events (corresponding to about 10 s of data taking) was 
generated using the HYDJET event generator \cite{hydjet} and processed through a full 
GEANT-4 detector simulation. This simulation includes realistic detector conditions and a 
full emulation of the Level-1 trigger system. The runtime performance of the HLT algorithms 
is evaluated by executing the trigger menu containing all trigger paths in a software framework 
equivalent to the online configuration used in the actual CMS HLT processor farm. 
Based on the full simulation of the detector response for a large scale data set a realistic 
timing estimate is achieved including all contributions of signal and physics background 
triggers. Especially in case of the dimuon trigger channel, which is seeded by the Level-1 
trigger objects, this strategy allows for a realistic estimate of the fraction of events 
that require the execution of the full dimuon reconstruction algorithm. Table~\ref{trigtime} shows the CPU time required to execute the individual trigger paths. The execution time is shown per minimum bias event and per module call. 

\begin{table}[hbt]
\begin{center}
\begin{tabular}{|l|c|c|c|}
\hline
Trigger Path & CPU time per event & CPU time per module call & Level-1 triggers per event\\ \hline
Min.bias              & $-$       &    $-$    &  $-$  \\ \hline
Jets           & 0.09 s      &    0.09 s   &   1 \\ \hline
Photons        & 0.14 s      &    0.14 s   &   1 \\ \hline
Dimuons        & 0.4 s       &    21 s     &   0.019 \\ \hline
\end{tabular}
\end{center}
\caption{\label{trigtime} CPU time required to execute the individual trigger paths.}
\end{table}

The full implementation of the CMS HLT processor farm for nominal running consists of about 10k CPU cores executing the event filter code. At the peak interaction rate of 8kHz (3kHz average) this corresponds to a CPU time budget of 1.25 (3.3) s available per event for reconstruction and to take a trigger decision. The total execution time for the full trigger menu was found to be $\approx$ 0.7 s per minimum bias Pb+Pb event and thus fits the available CPU budget including a sufficient safety margin. 
In Table~\ref{table_trigtable} an example trigger table is shown which illustrates 
the possible allocation of the total bandwidth to mass storage (225~MByte/s) to 
individual trigger channels. Clearly, this table will 
have to be optimized to maximize the scientific output of the CMS heavy ion program. 

\begin{table}[hbt]
\begin{center}
\begin{tabular}{|c|c|c|c|c|}
\hline
Channel & Threshold & Pre-scale & Bandwidth [MByte/s] & Event size [MB] \\ \hline
Min.bias              & $-$          &    1   &   33.75 (15\%)  & 2.5 \\ \hline
Jet                   & $100$~GeV    &    1   &   24.75 (11\%)  & 5.8 \\ \hline
Jet                   & $75$~GeV     &    3   &   27 (12\%)     & 5.7 \\ \hline
Jet                   & $50$~GeV     &   25   &   27 (12\%)     & 5.4 \\ \hline
Dimuons               &  $0$~GeV/c   &    1   &   69.75 (31\%)   & 4.9 \\ \hline 
Photons               & $10$~GeV     &    1   &   40.5 (18\%)   & 5.8 \\ \hline     
\end{tabular}
\end{center}
\caption{\label{table_trigtable} Example trigger table for heavy ion running at 
design luminosity, assigning fractions of the total bandwidth (225~MByte/s) to 
individual trigger channels. The last column shows the average event size for 
each of the trigger streams.}
\end{table}

Using this bandwidth allocation in the HLT, a gain of more than an order of magnitude is 
achieved for high \et~ jets and for dimuons compared to allocating the full bandwidth to 
writing minimum bias events to mass storage. A detailed discussion of the physics reach
achieved in heavy ion collisions using this trigger strategy can be found in \cite{gabor,hitdr}.

\section{Summary}

We have performed detailed simulations of the CMS HLT performance for studies of 
heavy ion collisions at the LHC. These studies validate our trigger strategy to 
only perform Pb+Pb event rejection in the HLT, based on the outcome of running 
reconstruction algorithms on the full event information.
This performance study is based on full event simulation and timing benchmarks 
from executing the heavy ion trigger menu in an online equivalent computing 
setup. The execution time of the full trigger menu of $\approx$ 0.7~s fits well within 
the available CPU time budget of 1.25 s given by the maximum interaction rate of 8kHz 
at nominal luminosity and the size of the HLT filter farm of about 10k CPU cores.
The proposed trigger menu allows to reduce the event rate given by the accelerator 
luminosity by prescaling abundant trigger signals while recording events passing the highest 
trigger threshold unprescaled. This trigger strategy allows to fully exploit 
the large production cross sections of hard probes at the LHC and, for example, enhances 
the recorded yield of high \et~ jets and dimuons by an order of magnitude compared 
to allocating the full bandwidth to minimum bias events.
This added statistical power provided by the HLT illustrates the crucial role 
the HLT plays for the CMS heavy ion physics program, which is essential for differential 
studies connecting rare probes to the physics of the QCD medium.



\end{document}